\documentclass[12pt,a4paper,leqno]{article}
\usepackage{graphicx}
\title{Statistical analysis of the water level of Huang He river (Yellow river) in China}
\author{Wang Bo$^1$, , Zlatinka I. Dimitrova$^2$, Nikolay K. 
Vitanov$^{3,4,}$\footnote{corresponding author, e-mail: vitanov@imbm.bas.bg}}
\date{$^1$ China-Central and Eastern Europe International Science and
Technology Achievement Transfer Center, Ningbo University of Technology,
201 Fenghua Rd., Jiangbei Qu, Ningbo Shi, Zhejiang Sheng 315000, China\\
$^2$ "G. Nadjakov" Institute of Solid State Physics, Bulgarian Academy of Sciences,
Blvd. Tzarigradsko Chaussee 72, 1784, Sofia, Bulgaria\\
$^3$ Institute of Mechanics, Bulgarian Academy of Sciences, Akad. G. Bonchev Str.,
Bl. 4, 1113 Sofia, Bulgaria\\
$^4$ Institute for Climate, Atmosphere and Water Research,  Bulgarian Academy
of Sciences, Blvd. Tsarigradsko Chaussee 66, 1784 Sofia, Bulgaria. }
\begin{document}
\maketitle
\begin{abstract}
Very high water levels of the large rivers are extremely dangerous events that can
lead to large floods and loss of property and  thousands and even tens of thousands 
human lives. The information from the  systematical
monitoring of the water levels allows us to obtain probability distributions for
the extremely high values of the water levels of the rivers of interest. In this article we study time series
containing information from more than 10 years of satellite observation
of the water level of the Huang He river (Yellow river) in China. We show that
the corresponding extreme values distribution is the Weibull distribution
and determine the parameters of the distribution. The obtained results may
help for evaluation of risks associated with floods  for the population and villages in the
corresponding region of the Huang He river. 
\end{abstract}
\begin{flushleft}
\end{flushleft}
\section{Introduction}
Complexity and nonlinearity of the dynamics are important features  of numerous
natural and social systems \cite{i1} - \cite{e1}. Several examples of models of such systems are the models
 of meteorology based on the Lorenz equations \cite{p1}, \cite{p2}, social dynamics models
 \cite{sd1} - \cite{sd5}, population dynamics models based on the Lotka–
Volterra equations and their generalizations \cite{v1}, turbulence theory \cite{v2} - \cite{v4}, 
theory of nonlinear waves \cite{rem} - \cite{sol2}, etc.. The model equations of the fluid mechanics systems
are nonlinear and thus the  mechanics of flows is a large source of nonlinear 
problems and models \cite{kam} -\cite{new2}. Because of this research in fluid mechanics is an area 
of extensive application of the methods of nonlinear dynamics and the theory of chaos \cite{ped} - \cite{new3}. 
In this article we shall apply the methodology of the theory of extreme values to a problem from the
research on the changing water levels of large rivers. As the number of floods increase in the last years 
the research  on the water levels intensifies as the large rivers can cause large losses of property and
human lives. In this research one uses the methods of the
nonlinear time series analysis and extreme values theory \cite{kantz} - \cite{reiss}.  These methods
\cite{ts1} -  \cite{v11} and their combination \cite{alb} are very effective for analysis of time series and for
description of many extreme events in Nature and society. 
\par
The global warming is another important reason for the  research interest ion water levels.
The global warming that leads to more rains and increasing of the levels of water that can cause floods and damages of
houses, cities, ports and people. This was the reason for the begin of our research on statistical description of
atmosphere and hydrological time series and study of the extreme values of these series \cite{v6},  \cite{h1}, \cite{h2}.
Below we shall study time series for the water level in for Huang He river in China. The Huang He (Yellow river) 
is the second longest river in Asia, after the Yangtze River, and the sixth longest river system in the world at the estimated length of 5\ 464 km.  The   total drainage area of the Huang He river is about 752,546 square kilometers. The Huang He
river was very important for the  ancient and contemporary  Chinese civilization. In the history and in the more recent times
frequent devastating floods and course changes of the flow of the river have been observed. These floods and changes are produced by the continual elevation of the river bed.  The waters of the Huang He river are responsible for large catastrophes including the only natural disasters in recorded history to have killed more than a million people: the flood from 1332 - 33  that killed 7 million people; the flood from  1887, which killed between 900,000 and 2 million people, and the flood from  1931 that 
killed between 1 and 4 million people. Because of all of  this the monitoring of the water level of the Huang He river and the study of the probabilities for extreme values of this level are important and even vital for the people of the regions along the river.
\par 
The organization of the text below is as follows. In Sect. 2 we describe the studied time series. In Sect. 3 we apply the
extreme values theory to the time series and calculate the parameters of the extreme value distributions for the water level of the Huang He river. Several concluding remarks are summarized in Sect. 4.
\section{The time series}
The studied time series - Fig. 1(a) are obtained from the x database \cite{dah}.
The water levels are measured on the base of satellite observation  of  the
Hung He river from 21.07.2008 till 02.02.2019 at longitude 115.1682$^o$ E and latitude
35.4222$^o$ N. The satellite measurement of the water level is performed one time per 10 days.
We observe significant changes in the water level of the river and the largest deviation from the average level
of the river  is more than two meters which is a large value and requires attention. 
\begin{figure}[tbh!]
	\vskip-2cm
	\hskip-1cm
	\includegraphics[scale=0.6,angle=-90]{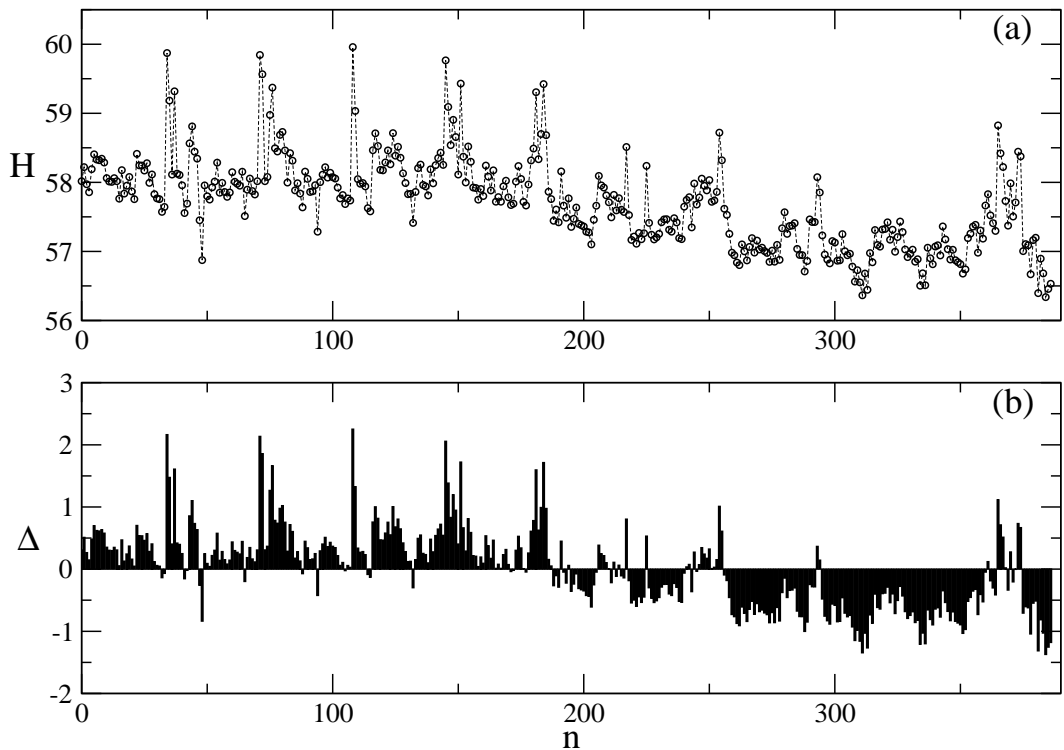}
	\caption{Time series of the water level of the Huang He river (Yellow river) (Figure (a)) and for 
		the deviation of the height of the water from its average value for the period of the records (Figure
		(b)). Horizontal axes: $n$ is the
		number of observation. Vertical axes: $H$ is the height of the water level in meters ($0$ is the 
		sea level); $\Delta$ is the deviation of the height of the water (in meters) from the average height of
		the water for the period of the records.}
\end{figure}
\par
The time series from Fig.1 show that in the period of measurement of more than 10 years the water
level of Huang He river had several  such larger deviations (of about two meters) from its average value. Another
interesting observation is  the decrease of the level of the river approximately in the middle of the
interval of the observation (from 2014). This decrease leaded to decreasing height of the maximum water levels: in the
first half of the observation period the maximum level was about 60 meters (height over the sea level). In the 
second part of the period of observation the maximum water level  does not exceed 59 meters.  
This decrease of the water level can be  seen also  in the relationship $\Delta(n)$
in Fig. 1(b). where at the right-hand side of the figure the water level in the larger part of the time is below
the average water level. We shall not discuss the reason for this drop of the water level. From the
point of view of the extreme values theory the drop of the water level will lead to changes of the parameters of the
extreme value distribution for the values of the water level. The the probability for extreme large water level (that can lead to
large floods) will change.
\begin{figure}[tbh!]
	\vskip-2cm
	\hskip-1cm
	\includegraphics[scale=0.6,angle=-90]{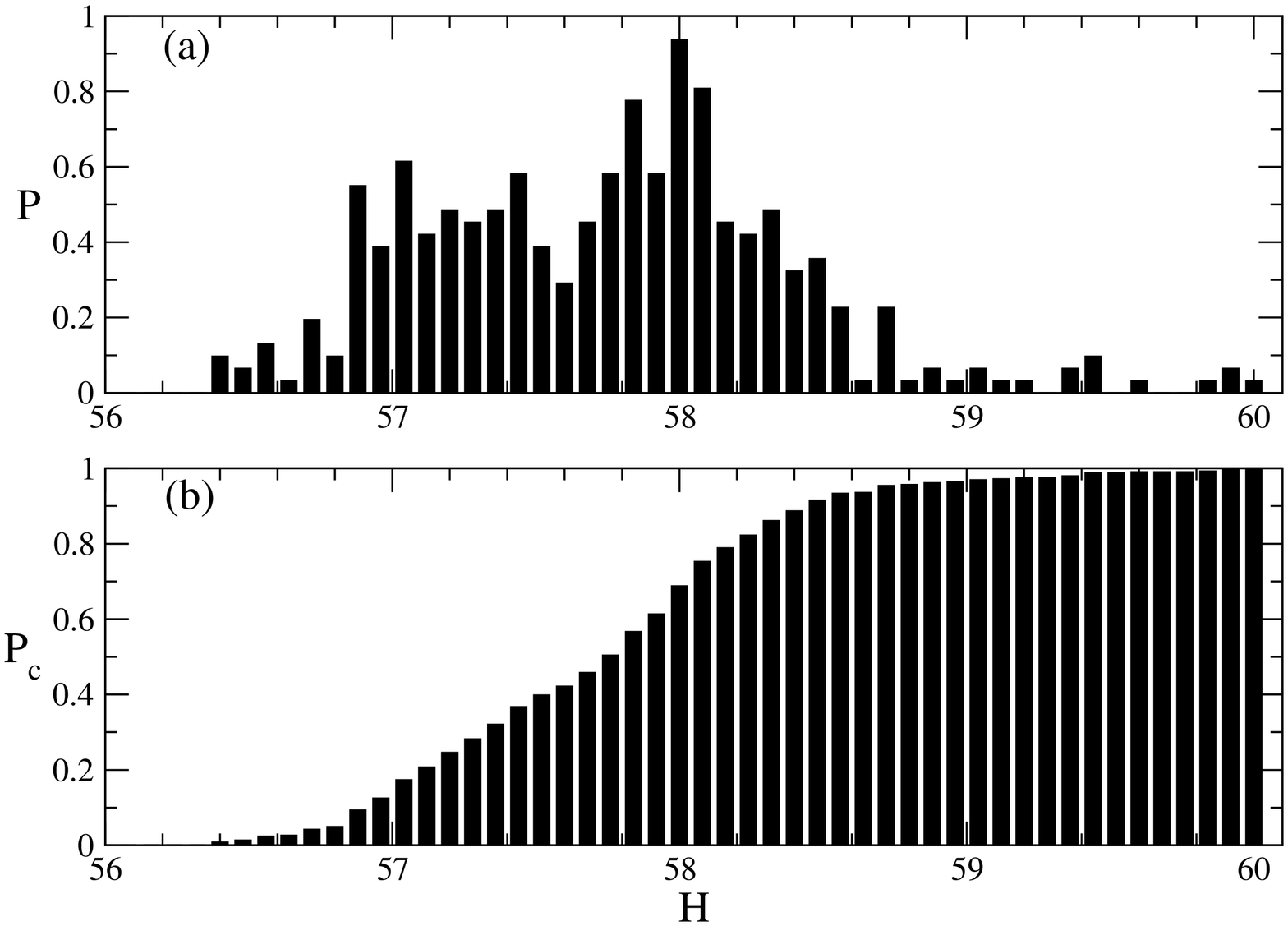}
	\caption{Histogram (Figure (a )) and cumulative histogram (Figure (b)) for the time series of 
		the water level of the Huang He river (Yellow river). Horizontal axes: $H$ is the
		height of the water level in meters ($0$ is the  sea level). Vertical axes: $P$: probability
		that the water level will be in the corresponding interval of values. $P_c$ : cumulative
		probability that the water level will be in the corresponding interval of values. $P_c$ }
\end{figure}
\par 
Fig. 2(a) shows the histogram and  Fig. 2(b) shows the cumulative histogram connected to the studied 
time series. In Fig. 2(a) we clearly observe presence of extreme values of the water level (far right-hand side of the figure).
 Another  interesting property of the histogram is presence of two areas of increased probability and area of decreasing probability (between $57.5$ and $57.7$ meters) of the water level of the Huang He river. This corresponds to the drop of
 the water level of the river. The cumulative histogram in Fig. 2(b) shows that the water level of the Huang He river
 newer dropped below $56.4$ m for the period of observation.
\section{Extreme value distributions}
The analyzed time series (Fig. 19a)) can be  divided into two approximately stationary parts
separated by a small region of non-stationary data. We shall consider the two approximately stationary regions.
We shall connect the water level of the river to a random variable.
An example for realization of a random variable are the temperatures at
a given place within a day. If we take the maximum daily temperature, we can
construct time series for the maximal daily temperatures. We shall treat in similar
way the time series for the water level. The water level rises and falls forming
a wave-like structure. We shall take the water level for every such structure as a
realization of a random variable and we shall investigate the time series for the
maxima of these variables, i.e. the time series for the maximum wave heights.
\par
Further we shall assume that the obtained maxima of the water level time series are maxima of a
stationary sequence of random variables $\{ H_k \}$, $k = 1, . . . , K$.  We are going to study
the limit distribution of these time series. In order to choose the appropriate methodology
for this study we have to account for the following important fact namely that  the limit distribution
does not need to be the same as the distribution of the maxima of the associated
independent sequence of random variables $\{ \tilde{H}̃_ k \}$, $k = 1, . . . , K$ with the same
marginal distribution as $\{ H_k \}$. Sometimes extreme values tend to occur in clusters and
we have to quantify this tendency for the studied time series. This can be made by 
 the extremal index $\theta$  which measures the tendency of the extreme values to occur in clusters. 
The mean size of the clusters of extreme values  is $1/\theta$.  Thus
if $\theta = 1$, the mean size of the clusters of extreme values is $1$ which means that there
is no clustering of the extreme events (we have a single extreme value and the values in the time series that
are around this extreme value are not extreme values). In this case
 the parameters of the extreme value distributions for the stationary
sequence $\{H_k \}$ and for the independent sequence $\{ \tilde{H}_k \}$ are the same.
\par
We shall consider two kinds of estimators for $\theta$. 
\begin{enumerate}
	\item \textit{The runs estimator}
\begin{equation}\label{runs}
\hat{\theta}_n^R(u,r) = \frac{\frac{1}{n-r}\sum \limits_{i=1}^{n-r} \textbf{1}(H_i>u, M_{i,i+r} \le u)}{\frac{1}{n} \sum \limits_{i=1}^n \textbf{1}(H_i > n)}
\end{equation}
where
\begin{itemize}
	\item   $H_i$, $i=1,\dots,n$ is the sample (e.g., the  time series which are  approximately stationary);
	\item $u$ is the threshold;
	\item  $r$ is the run length; 
	\item  $M_{ i,i+r}$ is the maximum value within the segment $H_{ i+1} , \dots ,H_{ i+r}$ ;
	\item \textbf{1} means that 1 is added to the corresponding sum if the condition in the () is fulfilled. 
	Otherwise 0 is added to the corresponding sum.
\end{itemize}
\item \textit{The interexceedance times estimator}
\begin{equation}\label{est2}
\hat{\theta}_N(u) = \frac{2 \left[ \sum \limits_{i=1}^{N-1} (T_i - 1)\right]^2}{(N-1) \sum \limits_{i=1}^{N-1} (T_i - 1) (T_i - 2)}
\end{equation}
where
\begin{itemize}
\item 
$N$ is the number of exceedances of the threshold value u and let these
exceedances happen at times $1 \le S_1 \le· · · \le S_N$ ;
\item
The interexceedance times are $T_ i = S_{i+1} - S_i$ .
\end{itemize}
We note that the interexceedance   estimator is appropriate when the maximum interexceedance time is larger than 2.
\end{enumerate}
\par
The application of the two estimators of $\theta$ to our time series shows that  for the
time series at the initial part of the period of observation (below the drop of the water level)
The $\theta$-index is  about $0.25$ up to threshold level of $57$ meters and then $\theta$ begins
to increase arriving at $\theta = 1$ (no clasterization of the extreme values) at threshold level of  $58.95$ meters.
For the case of the time series after the drop of the water level (right-hand side of the observed time series)
the $\theta$-index is about $0.25$ up to threshold level of $57.06$ meters (almost the same threshold as above)
and then the $\theta$-index begins to increase arriving at $\theta =1$ (no clustering of the extreme values) at
threshold level of $57.8$ meters.
\par 
\begin{figure}[tbh!]
	\vskip-2cm
	\hskip-1cm
	\includegraphics[scale=0.6,angle=-90]{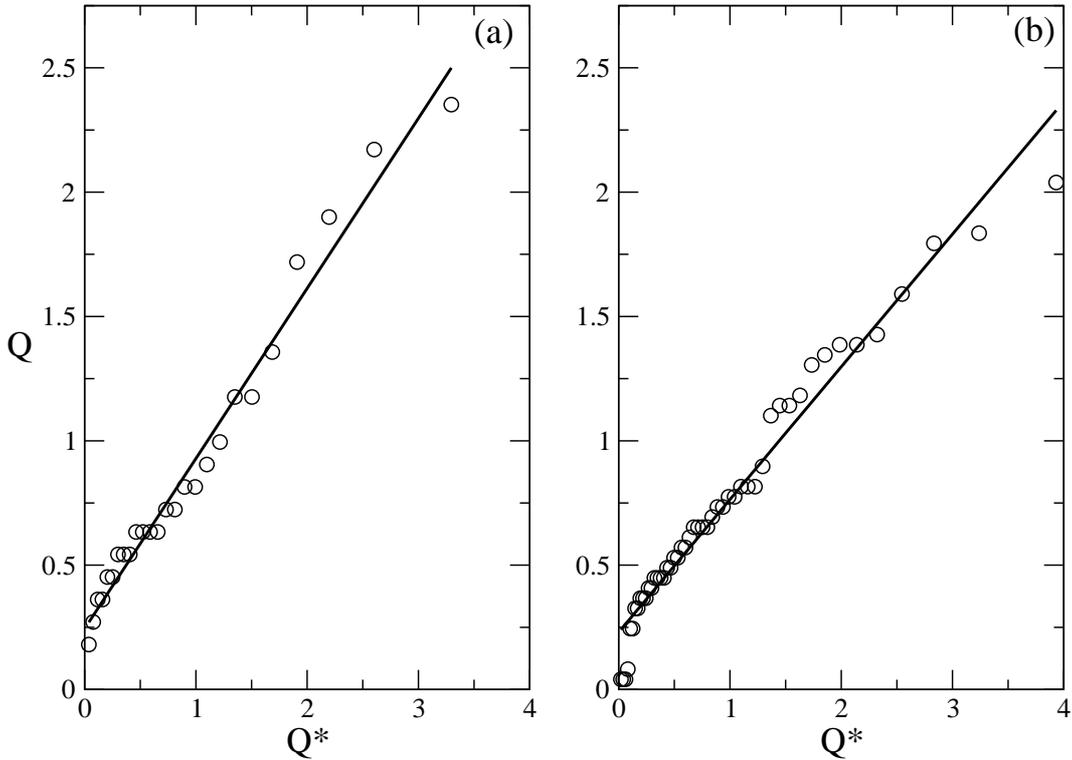}
	\caption{Quantile - quantile plot. $Q^*$: the standard exponential quantile. $Q$: quantiles for the
		water level of the Huang He river. Figure (a): quantile - quantile plot for the first half of the time series.
		Figure (b): quantile - quantile plot for the second half of the time series (after the drop of the water level). }
\end{figure}

Next we have to determine the kind of the distribution of the extreme values and to calculate the
parameters of this distribution. The kind of the distribution can be determined on the basis of quantile-quantile
plots. In order to make such plots we have to define the corresponding quantile function.
\par 
Let us have a time series $H_1 , H_2 , \dots , H_n$ with distribution function $F (x) =
P (H \le x)$. Important property of $F (x)$ is the inverse of the distribution function: the
quantile function $Q(p)$: 
$$Q(p) = \inf [x : F (x) \ge p].$$
We shall use the quantile functions much in the text below.
\par
On the basis of the series $H_i$ we can define the empirical distribution function
$\hat{F}(x) =\frac{i}{n}$ if $x \in [H_i , H_{i+1} ]$ and $H_i$ are ordered by increasing value. Then the 
empirical quantile function is
$$\hat{Q}(p) = \inf[x :\hat{F̂}(x) > p].$$
\par 
Let us use the quantile functions in order to produce some quantile - quantile plots for our data.
We consider first  the class of exponential distributions $$F_ \lambda(x) = 1 - \exp( - \lambda x)$$ where
$\lambda$ is a parameter. The quantile function for this class of  distributions is
$$
Q_\lambda(p) = - \frac{1}{\lambda} \ln (1-p)
$$
for $p \in (0, 1)$. For $\lambda = 1$ we have the standard exponential distribution and the
quantile of distribution with parameter $\lambda$ is connected to the standard exponential 
quantile function as follows $$Q_ \lambda (p) = \frac{1}{\lambda} Q_1 (p), \  \ p \in (0, 1).$$ 
Let us construct the standard exponential quantile-quantile $(Q-Q)$ plot: $[- \ln(1 - p), \hat{Q}(p)]$
where we plot the quantile of the exponential distribution vs. the quantile of the observed time series.
If the exponential distribution is a good fit to the experimental data $\hat{Q}(p)$ the standard
exponential quantile plot will be close to a straight line. The results for the water level time series of the
Huang He river are shown in Fig. 3.
We observe that the quantile - quantile plots are not close to a straight line for the case
of large values of the quantiles, i.e. for th case of large values of the water levels. This
means that the exponential distribution is not a good fit for the distribution of the
extreme values for the both time series for the water levels of the Huang He river.
\par 
\begin{figure}[tbh!]
	\vskip-2cm
	\hskip-1cm
	\includegraphics[scale=0.6,angle=-90]{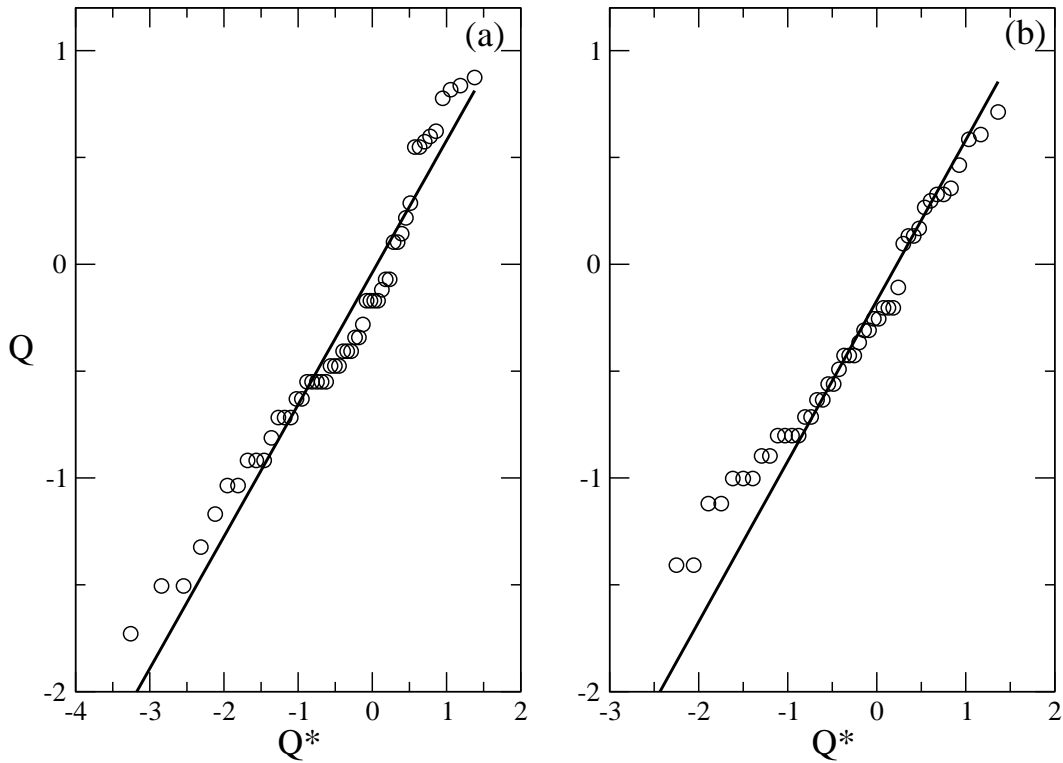}
	\caption{Quantile - quantile plot. $Q^*$: the quantile for the Weibull distribution. $Q$: quantiles for the
		water level of the Huang He river. Figure (a): quantile - quantile plot for the first half of the time series.
		Figure (b): quantile - quantile plot for the second half of the time series (after the drop of the water level). }
\end{figure}
The attempt of a direct fit shows that the exponential distribution is not a good fit for the distribution of
the extreme water levels of the Huang He river. The best fit exponential distribution is 
systematically lower than the values of the water levels for the case of large values of these levels.
Thus we have to search for another distribution that will fit the data  better.
\par
One such distribution can be a member of the class of the extreme values distributions.
There exist three kinds of extreme values distributions for i.i.d. random
variables: Weibull, Gumbel and Frechet-Pareto. If some of these distribution is
the extreme events distribution for the water level time series, the correspondent
quantile-quantile plot will be close to a straight line. We have obtained that 
that the Weibull distribution is the appropriate distribution for the maxima of
the water level time series. The Weibull distribution is
$$
F_W = 1 - \exp(-\lambda x^r ),
$$
where $ x > 0$, $\lambda > 0$ and $r > 0$. Probability density function for this distribution is
$f_ W = \lambda r x^{ r-1} \exp(-\lambda x^r )$. The quantile-quantile plot of Weibull quantile vs. the
quantile of the maxima of the water level time series is 
$$[\ln(- \ln(1 - p), \ln( \hat{Q}(p)))].$$
\begin{figure}[tbh!]
	\vskip-2cm
	\hskip-1cm
	\includegraphics[scale=0.6,angle=-90]{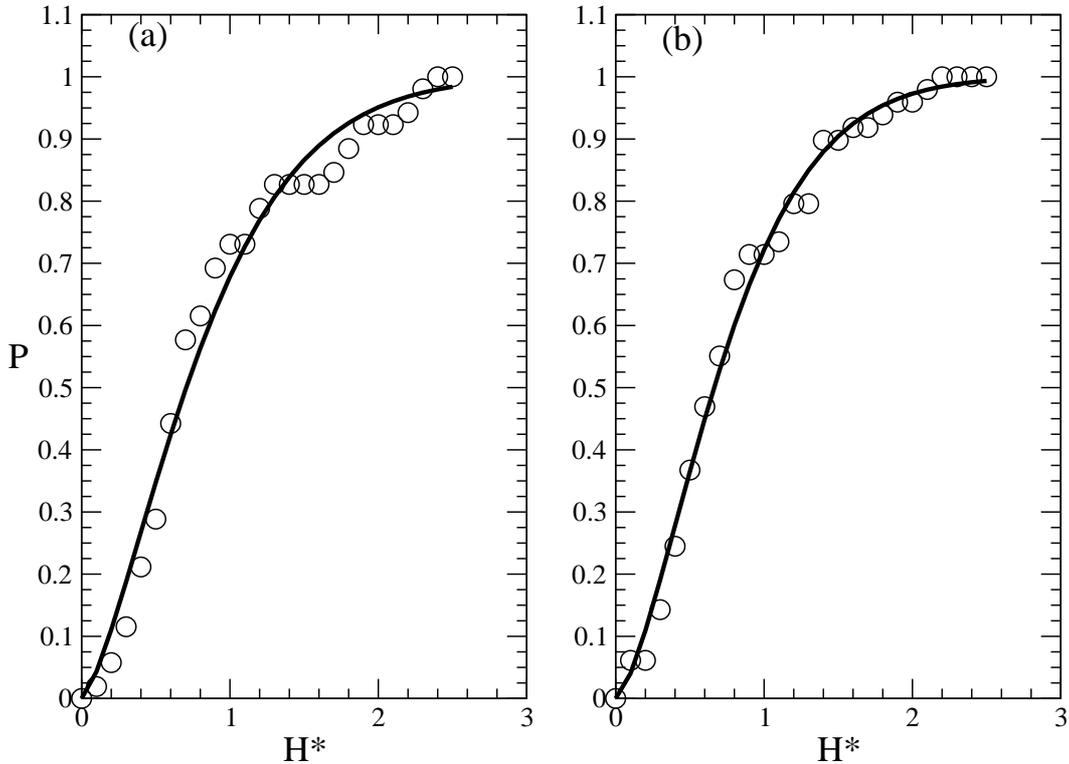}
	\caption{ The best fit of the extreme value distribution of the water levels of the Huang he river by the
		exponential distribution  $F_\lambda(x) = 1 - \exp(-\lambda x^r)$. $H^*$: relative heigh of the maximum over the threshold.
		$P$: probability. The values of the parameter $\lambda$ and $r$ are:
		Figure (a): $\lambda = 1.135$, $r=1.410$ ; Figure (b): $\lambda = 1.280$, $r=1.494$.}
\end{figure}
This plot is shown in Fig.4 for the two series of maximum values connected to the water level of the Huang He river.
We observe that the quantile-quantile plot is close to a straight line for large values of $H^*$ 
and this closeness is especially seen 
at the right-hand side of Fig. 4(b). Thus our next step is to fit the Weibull distribution to the distribution
of the maxima of the water level time series. The fit is shown in Fig. 5. This fit is much better in comparison
to the fit made on the basis of the exponential distribution. 
\par
The obtained extreme value distribution for the water level of the Huang He river allow us to calculate the
probabilities for very large values of these levels. These probabilities are summarized in Table 1.
\begin{table}[htb!]
	\centering
	\begin{tabular}{|c|c|c|}
\hline
Water level (m) & $P_1$	&  $P_2$    \\
\hline
\hline
   61  &    0.170                        & 0.00112    \\
\hline
	62	& 0.0104              & 0.0000176   \\
\hline
	63	&  0.000484         &  0.000000184  \\
\hline
	64	& 0.0000176        & 0.00000000132\\
\hline
	65  & 0.000000516   & 0.00000000000672\\
\hline
	66  &  0.0000000123& 0.0000000000000247\\  
\hline
	\end{tabular}
	\caption{Probability of corresponding large value of the water level of the Huang He river (in \%) before water level drop: $P_1$, and after the drop of the water level: $P_2$}
\end{table}
The results show that before the water drop one could expect water level of 61 m one time in about 5880 days (one time in 16 years) and after the drop of the water level this probability has decreased to 1 time in 89 285 days (one time in 245 years).
\section{Concluding remarks}
In this article we analyze  time series containing values for the water level 
of the second largest river in China - The Huang He river.
Our goals was to obtain the distribution for the extreme values of the water levels of the river around the point of the
measurement of the water level. We observe a drop of the water level of the river and divide the time series for the
water level into two parts: before the drop of the water level and after the drop of the water level.
We obtain that the extreme values distribution for the corresponding time series is the
Weibull distribution. We calculate the parameters of this distribution. On
the basis of the obtained distribution it is easy to obtain the probability of extreme
water levels. This may be of great help for the decision making about all the infrastructure, villages , town and cities
that can be affected by such high water levels. We note that results
are obtained on the basis of time series recorded in a given region of the river and the drop of the water level
leaded to significant decrease of the probability of extreme large water levels and corresponding large floods.

\end{document}